\documentclass[12pt]{article}
\pdfoutput=1
\usepackage[utf8]{inputenc}
\usepackage[DIV13]{typearea}
\usepackage{ragged2e}
\usepackage{amsmath,amsfonts,bbm,mathrsfs,amssymb}
\usepackage{slashed,cancel}
\usepackage[usenames,dvipsnames]{xcolor}
\usepackage{cite}
\usepackage{hyperref}
\hypersetup{colorlinks=true,urlcolor=Magenta,anchorcolor=blue,citecolor=blue,filecolor=blue,
            linkcolor=Magenta,menucolor=blue, linktocpage=true,pdfproducer=medialab}
\usepackage[english]{babel}
\usepackage{catchfile}
\usepackage{indentfirst}
\usepackage{graphicx}
\usepackage{float}
\usepackage{enumerate}
\usepackage{subcaption}
\usepackage{multirow}
\usepackage{tikz-feynman}
\usepackage[normalem]{ulem}

\newcommand{\email}[1]{\href{mailto:#1}{\tt #1}}

\numberwithin{equation}{section}

\newcommand{\blue}[1]{\color{blue} #1 \color{black}}

\definecolor{vierde}{rgb}{0.0, 0.5, 0.0}
\definecolor{OliveGreen}{rgb}{0,0.6,0}

\newcommand{\be}{\begin{equation}}
\newcommand{\ee}{\end{equation}}
\newcommand{\ba} {\begin{equation}\begin{aligned}}
\newcommand{\ea} {\end{aligned}\end{equation}}
\newcommand{\bea}{\begin{eqnarray}}
\newcommand{\eea}{\end{eqnarray}}

\newcommand{\YY}{\mathcal{Y}}
\newcommand{\LL}{\mathscr{L}}

\newcommand{\OO}{\mathcal{O}}
\newcommand{\cG}{\mathcal{G}}
\newcommand{\cM}{\mathcal{M}}

\newcommand{\unity}{\mathbbm{1}}

\newcommand{\derp}{\partial}
\newcommand{\hc}{\text{h.c.}}
\newcommand{\ov}[1]{\overline{#1}}

\newcommand{\mueV}{\ \mu\text{eV}}

\newcommand{\eV}{\ \text{eV}}

\newcommand{\TeV}{\ \text{TeV}}
\newcommand{\GeV}{\ \text{GeV}}
\newcommand{\MeV}{\ \text{MeV}}

\def\diag{{\tt diag}}
\def\Tr{{\rm Tr}}

\newcommand{\vev}[1]{\langle{#1}\rangle}

\begin{document}
\renewcommand*{\thefootnote}{\fnsymbol{footnote}}
\begin{titlepage}

\vspace*{-1cm}
\flushleft{FTUAM-20-16}
\hfill{IFT-UAM/CSIC-20-119}
\\[1cm]

\begin{center}
\bf\LARGE \blue{
Neutrino Masses and Hubble Tension}\\[2mm]
\bf\LARGE \blue{via a Majoron in MFV}\\[4mm]
\centering
\vskip .3cm
\end{center}
\vskip 0.5  cm
\begin{center}
{\large\bf Fernando Arias-Arag\'on}~\footnote{\email{fernando.arias@uam.es}},
{\large\bf Enrique Fern\'andez-Mart\'{\i}nez}~\footnote{\email{enrique.fernandez-martinez@uam.es}},\\[2mm]
{\large\bf Manuel Gonz\'alez-L\'opez}~\footnote{\email{manuel.gonzalezl@uam.es}}, and 
{\large\bf Luca Merlo}~\footnote{\email{luca.merlo@uam.es}}
\vskip .7cm
{\footnotesize
\centerline{ \it Instituto de F\'isica Te\'orica UAM/CSIC,}
\centerline{\it  Calle Nicol\'as Cabrera 13-15, Cantoblanco E-28049 Madrid, Spain}\vspace*{0.2cm}
\centerline{and}
\vspace*{0.2cm}
\centerline{\it Departamento  de  F\'{\i}sica Te\'{o}rica,  Universidad  Aut\'{o}noma  de  Madrid,}
\centerline{\it  Cantoblanco  E-28049  Madrid,  Spain}\vspace*{0.2cm}
}
\end{center}
\vskip 2cm
\begin{abstract}
\justify 
The recent tension between local and early measurements of the Hubble constant can be explained in a particle physics context. A mechanism is presented where this tension is alleviated due to the presence of a Majoron, arising from the spontaneous breaking of Lepton Number. The lightness of the active neutrinos is consistently explained. Moreover, this mechanism is shown to be embeddable in the Minimal (Lepton) Flavour Violating context, providing a correct description of fermion masses and mixings, and protecting the flavour sector from large deviations from the Standard Model predictions. A QCD axion is also present to solve the Strong CP problem. The Lepton Number and the Peccei-Quinn symmetries naturally arise in the Minimal (Lepton) Flavour Violating setup and their spontaneous breaking is due to the presence of two extra scalar singlets. The Majoron phenomenology is also studied in detail. Decays of the heavy neutrinos and the invisible Higgs decay provide the strongest constraints in the model parameter space.
\end{abstract}
\end{titlepage}
\setcounter{footnote}{0}

%\pdfbookmark[1]{Table of Contents}{tableofcontents}
%\tableofcontents

\renewcommand*{\thefootnote}{\arabic{footnote}}

%%%%%%%%%%%%%%%%%%%%%%%%%%%%%%%%%%%%%%%%%%%%%%%%%%%%%%%%%%%%
\section{Introduction}
%%%%%%%%%%%%%%%%%%%%%%%%%%%%%%%%%%%%%%%%%%%%%%%%%%%%%%%%%%%%
\label{sec:intro}

There is nowadays a considerable tension between late-time, local probes of the present rate of expansion of the Universe, that is the Hubble constant $H_0$, and its value inferred through the standard cosmological model $\Lambda$CDM from early Universe observations. Local measurements, from type-Ia supernovae and strong lensing, tend to cluster at similar values of $H_0$, significantly larger than those preferred by cosmic microwave background (CMB) and baryon acoustic oscillations probes. The strongest tension, estimated at the level of $4-6~\sigma$~\cite{Verde:2019ivm,Wong:2019kwg} depending on the assumptions performed, is between the {\tt Planck} inferred measure from the CMB spectrum~\cite{Aghanim:2018eyx} and the one obtained by the S$H_0$ES collaboration~\cite{Riess:2019cxk} from supernovae measurements.

Although the solution to this discrepancy might be related to systematics in the measurements (notably the callibration of the supernovae distances) or, more interestingly, point to a modification of the cosmological model, it may instead be provided by particle physics, as already discussed in the literature (see for example~\cite{Archidiacono:2016kkh,Ko:2016uft,DiValentino:2017oaw,DEramo:2018vss,Agrawal:2019lmo,Agrawal:2019dlm,Alexander:2019rsc,Ghosh:2019tab,Escudero:2019gzq,Escudero:2019gvw,Gelmini:2019deq,Park:2019ibn,Kreisch:2019yzn,Smith:2019ihp}). In particular, Ref.~\cite{Escudero:2019gvw} suggests that a Majoron that couples to light neutrinos could reduce the tension in the determination of $H_0$. It is then interesting to investigate whether this setup is compatible with possible explanations of other open problems in the Standard Model of particle physics (SM): the focus of this paper is to study the compatibility with Type-I Seesaw mechanism to explain the lightness of the active neutrinos, with specific flavour symmetries to describe the flavour puzzle and with the presence of an axion to solve the Strong CP problem. 

The Majoron, called $\omega$ hereafter,  is the Nambu-Goldstone boson (NGB) associated to the spontaneous breaking of lepton number (LN)~\cite{Chikashige:1980qk,Gelmini:1980re,Georgi:1981pg,Schechter:1981cv}, which is only accidental within the SM and breaks down at the quantum level. It naturally arises in the context of the Type-I Seesaw mechanism, where the Majorana mass term, instead of being a simple bilinear of the right-handed (RH) neutrino fields $N_R$, is a Yukawa-like term that couples $N_R$ to a scalar field that carries a LN charge, labelled as $\chi$ in the following. Once this scalar field develops a vacuum expectation value (VEV), then LN is spontaneously broken, a Majorana neutrino mass term is generated and the Majoron appears as a physical degree of freedom of the spectrum. 

If a Dirac term that mixes $N_R$ and the left-handed (LH) lepton doublets $\ell_L$ is also present in the Lagrangian, small masses for active neutrinos are generated at low energies, according to the Type-I Seesaw mechanism. 

At low-energies, the Majoron $\omega$ acquires a coupling with $\nu_L$, labelled as $\lambda_{\omega\nu\nu}$. For Majoron masses 
\be
m_\omega\in[0.1,\,1]\eV
\label{MajoronMassWindow}
\ee 
and $\lambda_{\omega\nu\nu}$ in the range 
\be
\lambda_{\omega\nu\nu}\in[5\times 10^{-14},\,10^{-12}]\,,
\label{MajoronNeutrinoMixingWindow}
\ee  
the tension on the Hubble constant is reduced~\cite{Escudero:2019gvw}. Indeed, for such small Majoron-neutrino mixings and Majoron masses, Majorons only partially thermalise after Big Bang Nucleosynthesis (BBN) or never thermalise~\cite{Chacko:2003dt}, enhancing the effective number of neutrino species $N_\text{eff}$ by at least $0.03$ and at most $0.11$, values that may be tested with CMB-S4 experiments~\cite{Abazajian:2016yjj}. Moreover, a non-vanishing $\lambda_{\omega\nu\nu}$ would reduce neutrino free-streaming, modifying the neutrino anisotropic stress energy tensor~\cite{Bashinsky:2003tk}. This has an impact in the CMB that results in modifying the posterior for the Hubble constant: as shown in Ref.~\cite{Escudero:2019gvw}, the inclusion of Majoron-neutrino interactions slightly shifts the central value of $H_0$, but largely broadens its profile reducing the $H_0$ tension to $2.5\sigma$. For larger couplings, $\lambda_{\omega\nu\nu}>10^{-12}$, these effects are too large and excluded by the same {\tt Planck} data.

Interestingly, Ref.~\cite{Escudero:2019gvw} found that the best $\chi^2$ in a Markov Chain Monte Carlo corresponds to Majoron mass and coupling as in Eqs.~\eqref{MajoronMassWindow} and \eqref{MajoronNeutrinoMixingWindow} and $\Delta N_\text{eff}=0.52\pm0.19$. The uncertainty in the last observable is very large and $\Delta N_\text{eff}=0$ is compatible within $3\sigma$. However, values close to the central one can be achieved if a thermal population of Majorons is produced in the early Universe and is not diluted during inflation. This may occur if the reheating temperature is larger than the RH neutrino masses~\cite{Escudero:2019gvw}. Alternatively, other relativistic species, such as axions~\cite{Masso:2002np,Salvio:2013iaa,Ferreira:2018vjj,Arias-Aragon:2020qtn}, may contribute to $\Delta N_\text{eff}$ and their presence may justify such a large value.

The existence of both Majorana and Dirac terms, the achievement of the correct scale for the active neutrino masses and at the same time the alleviation of the Hubble tension via the Majoron strongly depend on the LN charge assignments of $\ell_L$, $N_R$ and $\chi$. In particular, fixing the LN of $\ell_L$ to unity, then the LN of the RH neutrinos needs to have opposite sign with respect to the LN of the scalar field $\chi$. This model presents interesting phenomenological features. On one hand, the heavy neutrinos are expected to be relatively light, with masses at the MeV or GeV scales, opening up the possibility to be studied at colliders. Moreover, the presence of the Majoron may also have other consequences distinct from the Hubble tension.  In particular, its couplings to photons and electrons are constrained by CAST and Red Giant observations, respectively, while, due to its coupling to the Higgs, the Majoron may contribute to the invisible Higgs decay, strongly constrained by colliders. 

Sect.~\ref{sec:Mechanism} illustrates the mechanism to produce a Majoron that alleviates the $H_0$ tension together with a correct scale for the active neutrino masses. In Sect.~\ref{sec:Model}, this mechanism is introduced in a setup that correctly describes the flavour puzzle of the SM and at the same time produces a QCD axion that solves the Strong CP problem. Sect.\ref{sec:Signatures} gathers the phenomenological signatures of this model and Sect~\ref{sec:concl} contains the final remarks.

%%%%%%%%%%%%%%%%%%%%%%%%%%%%%%%%%%%%%%%%%%%%%%%%%%%%%%%%%%%%
\section{The Majoron Mechanism}
%%%%%%%%%%%%%%%%%%%%%%%%%%%%%%%%%%%%%%%%%%%%%%%%%%%%%%%%%%%%
\label{sec:Mechanism}

To produce a Majoron and explain the lightness of the active neutrinos, one can consider a Type-I Seesaw mechanism where the Majorana mass is dynamically generated by the spontaneous breaking of LN. The SM spectrum is extended by three RH neutrinos~\footnote{The case with only two RH neutrinos is also viable, and correspondingly the lightest active neutrino would be massless.} and a singlet scalar field $\chi$ that only transforms under LN. The LN charge assignments can be read in Tab.~\ref{tab:LN}, where $\ell_L$, $N_R$ and $\chi$ have already been defined, and $e_R$ refers to the RH charged leptons. Notice that  $L_\chi$ and $L_N$ are integer numbers and are not completely free, but must obey a series of constraints that will be made explicit in the following. 

\begin{table}[H]
\centering
\begin{tabular}{|c|c|}
\hline 
&\\[-4mm]
Field & $U(1)_L$ Charge \\[1mm] \hline\hline
&\\[-4mm]
$\ell_L$, $e_R$ & $1$ \\[1mm] \hline
&\\[-4mm]
$N_R$ & $-L_N$ \\[1mm] \hline
&\\[-4mm]
$\chi$ & $L_\chi$ \\[1mm] \hline
\end{tabular}
\caption{\em LN assignments. Fields that are not listed here do not transform under LN.}
\label{tab:LN}
\end{table}

The most general effective Lagrangian in the neutrino sector invariant under LN is the following~\footnote{Other terms can be added to this Lagrangian inserting $\chi^\dag$ instead of $\chi$. However, once the terms in Eq.~\eqref{Lag} are local, then their siblings with $\chi^\dag$ would not be local and therefore cannot be added to the Lagrangian. The only exception is the term
\be
\frac{1}{2}\left(\frac{\chi}{\Lambda_\chi}\right)^{\frac{2L_N+L_\chi}{L_\chi}} \chi^\dag \bar{N}^c_R\YY_NN_R\,,
\ee
that however only provides a suppressed correction with respect to the Majorana term written in Eq.~\eqref{Lag} and for this reason it can be neglected.}:
\begin{equation}
-\LL_{\nu} = \left(\frac{\chi}{\Lambda_\chi}\right)^{\frac{1+L_N}{L_\chi}}\bar{\ell}_L\tilde{H}\YY_\nu N_R + \frac{1}{2}\left(\frac{\chi}{\Lambda_\chi}\right)^{\frac{2L_N-L_\chi}{L_\chi}} \chi \bar{N}^c_R\YY_NN_R+\hc\,,
\label{Lag}
\end{equation}
where $H$ is the SM Higgs doublet, $\tilde{H}=i\sigma_2 H^*$, $\Lambda_\chi$ is the cut-off scale up to which the effective operator approach holds, and $\YY_\nu$ is a dimensionless and complex matrix, while $\YY_N$ is dimensionless, complex and symmetric. A first condition on $L_{N,\chi}$ arises from requiring that all the terms are local:
\be
\dfrac{1+L_N}{L_\chi}\,,
\dfrac{2L_N-L_\chi}{L_\chi} \in \mathbb{N}\,.
\label{FirstConditionxchi}
\ee

In the LN broken phase, the field $\chi$ can be parametrised as
\begin{equation}
\chi = \frac{\sigma+v_\chi}{\sqrt{2}}e^{i\frac{\omega}{v_\chi}},
\label{chiVEV}
\end{equation}
where the angular part $\omega$ is the NGB identified as a Majoron, $\sigma$ is the radial component and $v_\chi$ is its VEV. Notice that the scale appearing in the denominator of the exponent is also $v_\chi$ in order to obtain canonically normalised kinetic terms for the Majoron. A useful notation that will be employed in the following is the ratio of the $\chi$ VEV and the cut-off scale:
\be
\varepsilon_\chi=\frac{v_\chi}{\sqrt{2}\Lambda_\chi}.
\ee
This parameter $\varepsilon_\chi$ is expected to be smaller than 1 in order to guarantee a good expansion in terms of $1/\Lambda_\chi$. Consequently, the $\chi$ VEV, which represents the overall scale of the LN breaking, is expected to be smaller that the scale $\Lambda_\chi$, where New Physics should be present and is responsible for generating the expression in Eq.~\eqref{Lag}.

Once the electroweak symmetry is also spontaneously broken, i.e. after that the SM Higgs develops its VEV that in the unitary gauge reads
\be
H=\frac{h+v}{\sqrt{2}}\,,
\ee
where $h$ is the physical Higgs and $v\simeq246\GeV$, masses for the active neutrinos are generated according to the Type-I Seesaw mechanism:
\begin{equation}
\LL^\text{low-energy}_{\nu}=\dfrac{1}{2}\bar{\nu}_L\,m_\nu\,\nu^c_L +\hc
\qquad
\text{with}
\qquad
m_\nu=\frac{\varepsilon_\chi^{\frac{2+L_\chi}{L_\chi}} v^2}{\sqrt{2} v_\chi}\YY_\nu\YY_N^{-1}\YY_\nu^T\,.
\label{mass}
\end{equation}
In the basis where the charged lepton mass matrix is already diagonal, the neutrino mass matrix can be diagonalised by the PMNS matrix $U$:
\be
\hat{m}_\nu\equiv\diag\left(m_1,\,m_2,\,m_3\right)= U^\dagger\,m_\nu\,U^*\,.
\ee
The overall scale for the active neutrino masses can be written in terms of the parameter $\varepsilon_\chi$, the ratio of the VEVs and the product $\YY_\nu\YY_N^{-1}\YY_\nu^T$:
\be
\frac{\varepsilon_\chi^{\frac{2+L_\chi}{L_\chi}} v^2}{\sqrt{2} v_\chi}\YY_\nu\YY_N^{-1}\YY_\nu^T\simeq \sqrt{|\Delta m^2_\text{atm}|}\,,
\label{SecondConditionxchi}
\ee
where $\Delta m^2_\text{atm}=2.514^{+0.028}_{-0.027}\times 10^{-3}\eV^2$ for the Normal Ordering (NO) of the neutrino mass spectrum and $\Delta m^2_\text{atm}=-2.497\pm0.028\times 10^{-3}\eV^2$ for the Inverted Ordering (IO)~\cite{Esteban:2020cvm}.

The heavy neutrinos, that mostly coincide with the RH neutrinos, have a mass matrix that in first approximation can be directly read from the Lagrangian in Eq.~\eqref{Lag},  
\be
M_N\simeq\varepsilon_\chi^{\frac{2L_N-L_\chi}{L_\chi}} \frac{v_\chi}{\sqrt{2}}\mathcal{Y}_N\,.
\label{MN}
\ee
The overall scale for the heavy neutrinos must be larger than the overall scale of the active neutrinos, in order for the Seesaw approximation to hold:
\be
\varepsilon_\chi^{\frac{2L_N-L_\chi}{L_\chi}} \frac{v_\chi}{\sqrt{2}}\mathcal{Y}_N\gg\sqrt{|\Delta m^2_\text{atm}|}\,.
\label{ThirdConditionxchi}
\ee

On the other hand, the electroweak and LN breakings give rise to the Majoron Lagrangian that can be written as follows:
\begin{equation}
\begin{split}
\LL_\omega =& \frac{1}{2}\partial_\mu\omega\partial^\mu\omega +\frac{1}{2}m_\omega^2 \omega^2-
i\frac{1+L_N}{L_\chi}\varepsilon_\chi^{\frac{1+L_N}{L_\chi}}\frac{v}{\sqrt{2}v_\chi}\bar{\nu}_L\YY_\nu N_R \omega +
\\
&-i\frac{L_N}{L_\chi}\frac{\varepsilon_\chi^{\frac{2L_N-L_\chi}{L_\chi}}}{\sqrt{2}}\bar{N}^c_R\YY_NN_R \omega +\hc\ ,
\end{split}
\label{MajoronLag}
\end{equation}
where the $m_\omega^2$ term parametrises the Majoron mass introduced here as an explicit breaking of its corresponding shift symmetry. At low energy, a direct coupling of the Majoron to the active neutrinos emerges after performing the same transformations that give rise to the mass matrix in Eq.~\eqref{mass}:
\be
\LL^\text{low-energy}_\omega\supset i\frac{\lambda_{\omega\nu\nu}}{2}\omega\bar{\nu}_L\nu^c_L+\hc\footnote{This expression coincides with the one in Ref.~\cite{Escudero:2019gvw} once identifying $\omega$ with $\phi$ and $\lambda_{\omega\nu\nu}$ with $\lambda$.}
\qquad
\text{with}
\qquad
\lambda_{\omega\nu\nu}=2\dfrac{m_\nu}{L_\chi v_\chi}\,.
\label{lambdaomeganunu}
\ee
From the results in Ref.~\cite{Escudero:2019gvw}, shown in Eq.~\eqref{MajoronNeutrinoMixingWindow}, it is then possible to infer a bound on the product $L_\chi v_\chi$, once taking $\sqrt{\Delta m_\text{atm}^2}$ as the overall scale for the neutrino masses:
\be
|L_\chi| v_\chi\simeq\dfrac{2\sqrt{|\Delta m_\text{atm}^2|}}{\lambda_{\omega\nu\nu}}\in[0.1,\,2]\TeV\,.
\label{xchivchiRelation}
\ee
Adopting this relation and substituting it within the expressions in Eqs.~\eqref{SecondConditionxchi} and \eqref{ThirdConditionxchi}, new conditions can be found:
\begin{gather}
|L_\chi| \varepsilon_\chi^{\frac{2+L_\chi}{L_\chi}}\YY_\nu\YY_N^{-1}\YY_\nu^T
\simeq\dfrac{2\sqrt2}{\lambda_{\omega\nu\nu}}\dfrac{|\Delta m_\text{atm}^2|}{v^2}\in[1.2\times 10^{-13},\,2.4\times 10^{-12}]\,,
\label{SecondConditionxchiNew}\\
\dfrac{\varepsilon_\chi^{\frac{2L_N-L_\chi}{L_\chi}}}{|L_\chi|}\YY_N\gg\dfrac{\lambda_{\omega\nu\nu}}{\sqrt2}\simeq3.5\times 10^{-14}\,.
\label{ThirdConditionxchiNew}
\end{gather}

The following choice of charge assignments leads to a completely renormalisable Lagrangian and thus deserves special mention:

\be
\text{CASE R:}\qquad L_N=-1\qquad\text{and}\qquad L_\chi=-2\,,
\ee
such that the powers of the ratio $\chi/\Lambda_\chi$ in Eq.~\eqref{Lag} are not present in either the Dirac and the Majorana terms. The relation in Eq.~\eqref{xchivchiRelation} and the two conditions in Eqs.~\eqref{SecondConditionxchiNew} and \eqref{ThirdConditionxchiNew} further simplify:
\begin{align}
\text{Eq.~\eqref{xchivchiRelation}}\longrightarrow&\quad v_\chi\simeq\dfrac{\sqrt{|\Delta m_\text{atm}^2|}}{\lambda_{\omega\nu\nu}}\in[0.05,\,1]\TeV\,.
\label{xchivchiRelationR}\\
\text{Eq.~\eqref{SecondConditionxchiNew}}\longrightarrow&\quad\YY_\nu\YY_N^{-1}\YY_\nu^T
\simeq\dfrac{\sqrt2}{\lambda_{\omega\nu\nu}}\dfrac{|\Delta m_\text{atm}^2|}{v^2}\in[1.2\times 10^{-13},\,2.4\times 10^{-12}]\,,
\label{SecondConditionxchiNewR}\\
\text{Eq.~\eqref{ThirdConditionxchiNew}}\longrightarrow&\quad\YY_N\gg\dfrac{2\lambda_{\omega\nu\nu}}{\sqrt2}\simeq7\times 10^{-13}\,.
\label{ThirdConditionxchiNewR}
\end{align}
The first expression fixes a range of values for $v_\chi$. While the third expression represents a lower bound on the overall scale of $\YY_N$, the second one implies that the product $\YY_\nu\YY_N^{-1}\YY_\nu^T$ should be tuned to a very small value in order to reproduce the lightness of the active neutrinos.

For values of $L_{N,\chi}$ different from the previous ones, the Lagrangian is necessarily non-renormalisable. An interesting question is whether the extremely small values of the product $\YY_\nu\YY_N^{-1}\YY_\nu^T$ can be avoided exploiting the suppression in $\epsilon_\chi$ from the new physics scale $\Lambda_\chi$, similarly to the Froggatt-Nielsen approach to the flavour puzzle~\cite{Froggatt:1978nt}. Considering first the case in which $L_{N,\chi}>0$, then the only available possibilities are
\be
\begin{aligned}
\text{CASE NR1:}\qquad &L_N=1\qquad\text{and}\qquad L_\chi=1\\
\text{CASE NR2:}\qquad &L_N=1\qquad\text{and}\qquad L_\chi=2\,.
\end{aligned}
\ee
The corresponding values for $v_\chi$, $\varepsilon_\chi$, the overall scale for the heavy neutrinos $\langle M_N\rangle$ and the cut-off scale $\Lambda_\chi$ are reported in Tab.~\ref{tab:param}.
\begin{table}[H]
\hspace{-0.8cm}
\begin{tabular}{|c|c|c|c|c|c|c|}
\hline
&&&&&&\\[-4mm]
		& $L_N$ & $L_\chi$ & $v_\chi$ & $\varepsilon_\chi$ & $\langle M_N\rangle$ & $\Lambda_\chi$ \\[1mm] \hline
&&&&&&\\[-4mm]
CASE NR1 	& $1$ & $1$ & $[0.1,2]\TeV$ & $[0.49,1.4]\times 10^{-4}$ & $[3.5,200] \MeV$ & $[1.4-11]\times10^3\TeV$\\[1mm] \hline
&&&&&&\\[-4mm]
CASE NR2 	& $1$ & $2$ & $[0.05,1]\TeV$ & $[2.4,11]\times 10^{-7}$ & $[35.4,707] \GeV$ & $[1.4-6.5]\times10^5\TeV$\\[1mm]
\hline
\end{tabular}
\caption{\em Parameter ranges in the two phenomenologically interesting scenarios.}
\label{tab:param}
\end{table}

From Eq.~\eqref{SecondConditionxchiNew}, it can be seen that $\epsilon_\chi$ gets smaller for larger values of $L_\chi$ (unless tuning the product $\YY_\nu\YY_N^{-1}\YY_\nu^T$ as in CASE R discussed above, a possibility to be avoided in the present discussion): although this is not a problem by itself, it hardens the constraint in Eq.~\eqref{ThirdConditionxchiNew}. It follows that for larger values of $L_N$ and $L_\chi$ satisfying to the locality conditions in Eq.~\eqref{FirstConditionxchi}, then the overall scale of the heavy neutrino masses would be as small as the one of active neutrinos and therefore the expansion in the Type-I Seesaw mechanism would break down.

In the case when $L_N>0$ and $L_\chi<0$, it is possible to obtain the same results listed above substituting $\chi$ by $\chi^\dagger$ in the Lagrangian in Eq.~\eqref{Lag}: in this case, the signs in the denominators of the exponents would be flipped, compensating the negative sign of $L_\chi$. The opposite case, $L_N<0$ and $L_\chi>0$, is not allowed by the locality conditions.

For $L_{N,\chi}<0$, besides the possibility of CASE R, only another choice is allowed by the locality conditions:  $(L_N=-1, L_\chi=-1)$. However, this case would require $\epsilon_\chi \gg1$, leading to an even more extreme fine-tuning than in CASE R without the appeal of renormalizability.\\ 

The condition in Eq.~\eqref{xchivchiRelation}, corresponding to Eq.~\eqref{MajoronNeutrinoMixingWindow}, is only one of the ingredients necessary to lower the $H_0$ tension. A second relevant requirement is Eq.~\eqref{MajoronMassWindow}, regarding the Majoron mass. For the sake of simplicity, it has been introduced directly in the Majoron Lagrangian in Eq.~\eqref{MajoronLag} as en effective term. Its origin has been widely discussed in the literature and constitutes in itself an interesting research topic. Any violation of the global LN symmetry would induce a mass for the Majoron. An obvious example are gravitational effects, which are expected to break all accidental global symmetries. Estimations of their size from non-perturbative arguments via wormhole effects \cite{Alonso:2017avz} fall too short of their required value. Conversely, their size from Planck-suppressed effective operators~\cite{Akhmedov:1992hi} would point to too large a mass, although several possibilities have been discussed that would prevent the lower dimension operators from being generated~\cite{Dobrescu:1996jp,Lillard:2018fdt,Hook:2019qoh,Dienes:1999gw,Choi:2003wr,Cox:2019rro,Fukuda:2017ylt,Carone:2020nlx}. These options were originally introduced as a solution to the axion quality problem. A simpler possibility, given the singlet nature of the $N_R$, is an explicit breaking of LN via a Majorana mass term at the Lagrangian level. The Majoron would thus develop a mass slightly below this breaking scale from its coupling to the $N_R$ through self-energy diagrams. In this work we will remain agnostic to the origin of the Majoron mass, and a value consistent with Eq.~\eqref{MajoronMassWindow} will be assumed. \\

The mechanism illustrated in this section allows to soften the $H_0$ tension explaining at the same time the lightness of the active neutrinos. In the next section, this mechanism will be embedded into a specific flavour model that allows to account for the Flavour Puzzle, without violating any bounds from flavour observables, and also contains a QCD axion that solves the Strong CP Problem.

%%%%%%%%%%%%%%%%%%%%%%%%%%%%%%%%%%%%%%%%%%%%%%%%%%%%%%%%%%%%
\section{The Majoron and Axion from an MFV Setup}
%%%%%%%%%%%%%%%%%%%%%%%%%%%%%%%%%%%%%%%%%%%%%%%%%%%%%%%%%%%%
\label{sec:Model}

Flavour models aim at describing the heterogeneity of the fermion masses and mixings through some underlying argument, such as a symmetry principle. The simplest proposal is known as the Froggatt-Nielsen model~\cite{Froggatt:1978nt} and consists in introducing a global Abelian symmetry to describe the flavour structure in the quark sector. Almost 25 years later, after the more precise measurements of neutrino oscillations and the introduction of the so-called Tri-Bimaximal mixing texture~\cite{Harrison:2002er,Xing:2002sw}, whose main feature is a vanishing reactor mixing angle in the PMNS matrix, discrete symmetries were considered as an attractive approach to reproduce the observed pattern of all fermion masses and mixings~\cite{Ma:2001dn,Babu:2002dz,Altarelli:2005yp,Ishimori:2010au,Altarelli:2012ss,Hernandez:2012ra,Grimus:2011fk,King:2013eh}. However, after the discovery, in 2011, of a relatively large reactor mixing angle~\cite{Abe:2011sj,Adamson:2011qu,Abe:2011fz,An:2012eh,Ahn:2012nd}, models based on discrete symmetries underwent a deep rethinking and other options also became popular. A few examples are Froggatt-Nielsen inspired models based on a simple $U(1)$~\cite{Buchmuller:2011tm,Altarelli:2012ia,Bergstrom:2014owa} and models based on continuous non-Abelian symmetries. The latter include Minimal Flavour Violation (MFV)~\cite{DAmbrosio:2002vsn} and its leptonic versions~\cite{Cirigliano:2005ck,Davidson:2006bd,Alonso:2011jd}, which are based on a $U(3)$ symmetry, smaller symmetries like $U(2)$~\cite{Barbieri:2011ci,Blankenburg:2012nx}, or an intermediate approach~\cite{Arias-Aragon:2020bzy}.

The focus here will be the MFV setup, that will be shown to naturally suggest the presence of the Majoron and of a QCD axion. The key concept behind MFV is to require that any flavour and CP violation in physics Beyond the SM has the same origin as the one in the SM~\cite{Chivukula:1987py}. This idea has been formulated in terms of the symmetries of the kinetic terms in Ref.~\cite{DAmbrosio:2002vsn}. These symmetries are then broken by a series of fields that are also employed to describe fermion masses and mixings, as well as the suppressions associated to any non-renormalisable flavour violating operator. Considering the SM spectrum augmented by three RH neutrinos, the flavour symmetry of the corresponding kinetic terms is a product of a $U(3)$ term for each fermion species,
\be
\cG_F=U(3)^6\,.
\ee
The non-Abelian terms are responsible for the intergenerational fermion mass hierarchies and of the mixing matrices, while the Abelian terms are associated to the hierarchies among the masses of the different third family fermions. The latter can be written as follows
\begin{equation}
\cG_F^A=U(1)_{q_L}\times U(1)_{u_R}\times U(1)_{d_R}\times U(1)_{\ell_L}\times U(1)_{e_R} \times U(1)_{N_R}\,,
\end{equation}
where $q_L$ refers to the LH quark doublets, $u_R$ to the RH up-type quarks and $d_R$ to the RH down-type quarks, while $\ell_L$, $e_R$ and $N_R$ to the leptons as already introduced in the previous section. It is possible to rearrange this product, provided that the new combinations are still linearly independent, identifying among them baryon number, LN, weak hypercharge and the Peccei-Quinn (PQ)~\cite{Peccei:1977hh} symmetry:
\be
\cG_F^A=U(1)_B\times U(1)_L \times U(1)_Y \times U(1)_{PQ} \times U(1)_{e_R} \times U(1)_{N_R}\,.
\label{U1symmetries}
\ee
In the model described in this section, fermion charges under baryon number and hypercharge are assigned as in the SM, while the LN charges are given in Tab.~\ref{tab:LN}. Moreover, PQ charges are chosen so as to explain the suppression of the bottom and $\tau$ masses with respect to the top mass: only $d_R$ and $e_R$ transform under PQ, with a charge equal to 3. In an analogous way to LN, the PQ symmetry is formally exact at the Lagrangian level after introducing a second scalar field $\Phi$ that transforms non-trivially only under PQ with a charge $-1$:
\begin{equation}
\Phi=\frac{\rho+v_\Phi}{\sqrt{2}}e^{i\frac{a}{v_\Phi}},
\end{equation}
where $\rho$ is the radial component, $a$ is the angular one and $v_\Phi$ is its VEV. Once this scalar field develops a VEV and electroweak symmetry gets broken, masses for bottom and $\tau$ are generated, suppressed with respect to the top one. As a byproduct of this mechanism, the axion $a$ is originated~\cite{Arias-Aragon:2017eww}. 
Finally, the last two symmetries in Eq.~\eqref{U1symmetries} do not play any role in this model and are explicitly broken. 

The Lagrangian invariant under the aforementioned symmetries is the following:
\begin{equation}
\begin{split}
-\LL_{Y}=&\bar{q}_L\tilde{H}\YY_uu_R+\left(\frac{\Phi}{\Lambda_\Phi}\right)^3\bar{q}_L{H}\YY_dd_R+\left(\frac{\Phi}{\Lambda_\Phi}\right)^3\bar{\ell}_L{H}\YY_ee_R+\\
&+\left(\frac{\chi}{\Lambda_\chi}\right)^{\frac{1+x_N}{x_\chi}}\bar{\ell}_L\tilde{H}\YY_\nu N_R+\frac{1}{2}\left(\frac{\chi}{\Lambda_\chi}\right)^{\frac{2x_N-x_\chi}{x_\chi}} \chi \bar{N}^c_R\YY_NN_R+\hc\,,
\end{split}
\end{equation}
where $\Lambda_\Phi$ stands for the cut-off scale associated to the scalar field $\Phi$. 

According to the MFV approach,  $\YY_{u,d,e,\nu,N}$ are not simple matrices, but are promoted to be spurion fields that do transform under the non-Abelian part of the flavour symmetry group $\cG_F$. These fields can be thought of as dimensionless scalar fields that  do not have kinetic terms, but acquire background values that play the role of VEVs. In the MFV approach, to reproduce correctly quark masses and mixings and charged lepton masses, the background values of $\YY_{u,d,e}$ should read
\be
\begin{aligned}
\left\langle \mathcal{Y}_u \right\rangle & = c_t\,V^\dagger\, \diag\left(\frac{m_u}{m_t},\frac{m_c}{m_t},1 \right) \,,\\
\left\langle \mathcal{Y}_d \right\rangle & = c_b\,\diag\left(\frac{m_d}{m_b},\frac{m_s}{m_b},1 \right) \,,\\
\left\langle \mathcal{Y}_e \right\rangle & = c_\tau\, \diag\left(\frac{m_e}{m_\tau},\frac{m_\mu}{m_\tau},1 \right) \,,
\end{aligned}
\ee
where $V$ is the CKM mixing matrix and $c_i$ are order 1 parameters. The possible origin of these values is under study~\cite{Alonso:2011yg,Alonso:2012fy,Alonso:2013mca,Alonso:2013nca}. 

While these matrices present hierarchies among their entries, the ratios $m_b/m_t$ and $m_\tau/m_t$ are still not described within this approach, but can be explained by the spontaneous breaking of the PQ symmetry. Indeed, the overall scale of down-type quarks and charged leptons is multiplied by the cubic power of the ratio among the $\Phi$ VEV and the cut-off $\Lambda_\Phi$. Assuming that this ratio is given by
\be
%\varepsilon_\Phi\equiv
\dfrac{v_\Phi}{\sqrt{2}\Lambda_\Phi}\simeq0.23\,,
\ee
the $m_b/m_t$ and $m_\tau/m_t$ ratios are also correctly described.

Any non-renormalisable operator that describes flavour-violating processes should be invariant under the flavour symmetry. This is accomplished by inserting proper powers of the spurions: once they acquire their background values, the non-renormalisable operator under consideration gets suppressed. The main consequence is that the scale that can be studied considering flavour observables is at the level of $1-10 \TeV$~\cite{DAmbrosio:2002vsn,Grinstein:2006cg,Grinstein:2010ve,Feldmann:2010yp,Guadagnoli:2011id,Redi:2011zi,Buras:2011zb,Buras:2011wi,Alonso:2012jc,Alonso:2012pz,Lopez-Honorez:2013wla,Merlo:2018rin}, instead of $100\TeV$ in a generic case~\cite{Isidori:2010kg}, opening up the possibility of discovering New Physics (NP) at colliders.

For the neutrino sector, the discussion is slightly more complicated. Indeed, there are two spurions in the neutrino sector, $\YY_\nu$ and $\YY_N$, and both enter in the definition of the active neutrino masses, see Eq.~\eqref{mass}. It follows that it is not possible to identify univocally either $\YY_\nu$ or $\YY_N$ in terms of neutrino masses and the PMNS matrix entries. Thus, the suppression in the non-renormalisable flavour violating operators cannot be directly linked to neutrino masses and mixings, losing the predictivity that characterizes the MFV approach in the quark sector. The solutions that have been proposed are to assume $\YY_N\propto\unity$~\cite{Cirigliano:2005ck,Davidson:2006bd} or to consider $\YY_\nu$ as a unitary matrix~\cite{Alonso:2011jd}. 

\begin{itemize}
\item[I):] $\cG^{NA}_L\to SU(3)_{\ell_L}\times SU(3)_{e_R}\times SO(3)_{N_R}\times CP$~\cite{Cirigliano:2005ck,Davidson:2006bd}.\\[5pt]
Under the assumption that the three RH neutrinos are degenerate in mass, i.e. \mbox{$\YY_N\propto \unity$}, then the non-Abelian symmetry associated to the RH neutrinos, $SU(3)_{N_R}$, is broken down to $SO(3)_{N_R}$. The additional assumption of no CP violation in the lepton sector is meant to force $Y_\nu$ to be real\footnote{Strictly speaking, the condition of CP conservation in the leptonic sector forces the Dirac CP phase to be equal to $\delta^\ell_{CP}=\{0,\,\pi\}$ and the Majorana CP phases to be $\alpha_{21,31}=\{0,\,\pi,\,2\pi\}$. However, $Y_\nu$ is real only if $\alpha_{21,31}=\{0,\,2\pi\}$, and therefore $\alpha_{21,31}=\pi$ needs to be disregarded in order to guarantee predictivity. The CP conservation condition assumed in Refs.~\cite{Cirigliano:2005ck,Davidson:2006bd} is then stronger than the strict definition.}. With these simplifications, the expression for the active neutrino mass in Eq.~\eqref{mass} simplifies to
\be
m_\nu=\frac{\varepsilon_\chi^{\frac{2+L_\chi}{L_\chi}} v^2}{\sqrt{2} v_\chi}\YY_\nu\YY_\nu^T\,,
\label{massMLFVI}
\ee
and all flavour changing effects involving leptons can be written in terms of $\YY_\nu \YY_\nu^T$ and $Y_e$. These are the only relevant combinations entering any $d=6$ operator involving lepton fields and describing flavour violating effects. It follows that any flavour changing process can be predicted in terms of lepton masses and mixings.

Diagonalising $m_\nu$ corresponds to diagonalising the product $\vev{\YY_\nu} \vev{\YY_\nu}^T$ and, given the fact that the the lepton mixing angles are relatively large, then no hierarchies should be expected among the entries of $\vev{\YY_\nu}\vev{ \YY_\nu}^T$, contrary to what happens in the quark case. Note that some setups, such as the so-called sequential dominance scenarios, obtain large mixing angles even if there exists a strong hierarchy among the Yukawa couplings~\cite{King:1999mb}. However, this possibility is disfavoured by the general philosophy of MLFV. In the same spirit, the overall scale of this product is of $\OO(1)$, as any explanation of the neutrino masses should reside in the model itself, and not be due to any fine-tuning.

\item[II):] $\cG^{NA}_L\to SU(3)_{\ell_L+N_R}\times SU(3)_{e_R}$~\cite{Alonso:2011jd}.\\[5pt]
Assuming that the three RH neutrinos transform as a triplet under the same symmetry group of the lepton doublets, 
\be
\ell_L, N_R\sim ({\bf 3},\,1)_{\cG^{NA}_L} \qquad\qquad
e_R\sim (1,\,{\bf 3})_{\cG^{NA}_L}\,,
\ee
then Schur's Lemma guarantees that $\YY_\nu$ transforms as a singlet of the symmetry group. Then, $Y_\nu$ is a unitary matrix~\cite{Bertuzzo:2009im,AristizabalSierra:2009ex}, which can always be rotated to the identity matrix by a suitable unitary transformation acting only on the RH neutrinos. The only meaningful quantities in this context are $\YY_e$ and $\YY_N$, so neutrino masses and lepton mixings are encoded uniquely into $Y_N$,
\be
m_\nu=\frac{\varepsilon_\chi^{\frac{2+L_\chi}{L_\chi}} v^2}{\sqrt{2} v_\chi}\YY_N^{-1}\,.
\label{massMLFVII}
\ee
Moreover, all flavour changing effects involving leptons can be written only in terms of $Y_e$ and $Y_N$,
and therefore any flavour changing process can be predicted in terms of lepton masses and mixings. 

As for the previous case, the diagonalisation of the active neutrino mass coincides with the diagonalisation of $\vev{\YY_N}^{-1}$, that therefore does not present any strong hierarchy among its entries. Moreover, its overall scale should be $\OO(1)$ according to the MLFV construction approach. 
\end{itemize}

In both cases, the constraints on NP considering the present available data on flavour changing processes in the lepton sector are as low as a few TeV~\cite{Cirigliano:2005ck,Cirigliano:2006su,Davidson:2006bd,Gavela:2009cd,Alonso:2011jd,Alonso:2016onw,Dinh:2017smk}.\\

Once the PQ symmetry is spontaneously broken, the axion arises as its NGB and its Lagrangian can be written as
\be
\begin{aligned}
\LL_a=&\dfrac{1}{2}\derp_\mu a\derp^\mu a-e^{\frac{3ia}{v_\Phi}}\ov{q}_LH\YY_d d_R-e^{\frac{3ia}{v_\Phi}}\ov{\ell}_LH\YY_e e_R+\dfrac{\alpha_s}{8\pi}\theta_{\mathrm{QCD}}G^{a\mu\nu}\widetilde{G}^a_{\mu\nu}\,,
\label{InitialLagrangianLE}
\end{aligned}
\ee
with $\widetilde{G}^a_{\mu\nu}\equiv \frac{1}{2}\epsilon_{\mu\nu\rho\sigma} G^{a\rho\sigma}$ and $\epsilon_{\mu\nu\rho\sigma}$ the totally antisymmetric tensor such that $\epsilon_{1230}=1$. The last term is the well-known QCD $\theta$-term, allowed by the QCD Lagrangian, which constitutes a source of CP violation. The $\theta_{\mathrm{QCD}}$ parameter contributes to the neutron electric dipole moment~\cite{Baluni:1978rf,Crewther:1979pi} and can thus be experimentally constrained~\cite{Abel:2020gbr}
\be
\theta_{\mathrm{QCD}}\lesssim10^{-10}\,.
\ee
The presence of the axion naturally explains why $\theta_{\mathrm{QCD}}$ is so small, providing a solution to the Strong CP problem: couplings of the axion to the gauge fields, and in particular to gluons, are generated at the quantum level and the $\theta_{\mathrm{QCD}}$ parameter can be reabsorbed by an axion field redefinition~\cite{Peccei:1977hh,Wilczek:1977pj,Weinberg:1977ma}. The effective axion potential (see for example Ref.~\cite{diCortona:2015ldu}) predicts a vanishing VEV for the axion, that finally solves the Strong CP problem. Moreover, the same potential provides the axion with a mass,
\be
m_a\sim 6\mueV\left(\dfrac{10^{12}\GeV}{f_a}\right)\,,
\ee
being $f_a$ the axion scale, that is connected to the $\Phi$ VEV in this model by the relation $f_a=v_\Phi/9$.
For QCD axions, the most stringent constraint on $f_a$ comes from the axion couplings to photons~\cite{Jaeckel:2015jla,Bauer:2017ris,Anastassopoulos:2017ftl} and to electrons~\cite{Viaux:2013lha,Straniero:2018fbv,Diaz:2019kim}, which push its value to be larger than $f_a\gtrsim1.2\times 10^7\GeV$ and $f_a\gtrsim8\times 10^8\GeV$, respectively.\\

Summarising, the Majoron together with axion constitute the natural Abelian completion of MFV scenarios. The Majoron does not affect (at tree level) the physics associated to the axion and the quark and charged lepton flavour physics. Thus, this model, besides describing fermion masses and mixings and solving the Strong CP problem, is able to alleviate the Hubble tension with the only inclusion of three RH neutrinos and two extra singlet scalars. In particular, as no fine-tuning is allowed within this approach on $\vev{\YY_\nu}$ or $\vev{\YY_N}$, then only CASE NR1 and CASE NR2 are viable in the MFV framework. In the following section, the analysis of this model will be completed with the study of its phenomenological signatures.

%%%%%%%%%%%%%%%%%%%%%%%%%%%%%%%%%%%%%%%%%%%%%%%%%%%%%%%%%%%%
\section{Phenomenological Signatures}
%%%%%%%%%%%%%%%%%%%%%%%%%%%%%%%%%%%%%%%%%%%%%%%%%%%%%%%%%%%%
\label{sec:Signatures}

The only tree-level coupling of the Majoron is to neutrinos. However, at quantum level, couplings to gauge bosons, other SM fermions and the Higgs are originated. 

\subsection*{Coupling to photons}
The searches for very light pseudoscalars, usually addressed to axions, can also apply to Majorons. In the range of masses in Eq.~\eqref{MajoronMassWindow}, the strongest constraints on the effective coupling to photons are set by CAST~\cite{Anastassopoulos:2017ftl}, which establishes the upper bound 
\be
\lambda_{\omega\gamma\gamma}\lesssim 10^{-10}\GeV^{-1}\,,
\ee
where $\lambda_{\omega\gamma\gamma}$ is defined as
\be
\LL_\omega^\text{low-energy} \supset \frac{1}{4}\,\lambda_{\omega\gamma\gamma}\,\omega\, F^{\mu\nu} \tilde{F}_{\mu\nu}
\label{EffectiveOmega2Photons}
\ee
with $\widetilde{F}_{\mu\nu}\equiv \frac{1}{2}\epsilon_{\mu\nu\rho\sigma} F^{\rho\sigma}$.

As the Majoron does not couple at tree-level to charged particles, then the process $\omega\to\gamma\gamma$ occurs only at two loops. Ref.~\cite{Garcia-Cely:2017oco} provides an estimate for its decay width: under the assumption $m_\omega \ll m_e$,
\begin{equation}
\Gamma_{\omega\to \gamma\gamma}=\frac{\alpha^2}{1536^2\pi^7}\dfrac{m_\omega^7}{v^2m_e^4}\left(\Tr\left[\dfrac{m_Dm_D^\dagger}{vv_\chi}\right]\right)^2
\end{equation}
where $\alpha\equiv e^2/4\pi$ and with $m_D$ the Dirac neutrino mass matrix. 

Computing the same process by means of the effective couplings in Eq.~\eqref{EffectiveOmega2Photons}, 
\be
\Gamma_{\omega\to \gamma\gamma}=\dfrac{\lambda_{\omega\gamma\gamma}^2m_\omega^3}{32\pi}\,
\ee
it is then possible to match the two expressions for the $\omega\to\gamma\gamma$ decay width providing the expression for the $\lambda_{\omega\gamma\gamma}$ coupling:
\begin{equation}
\lambda_{\omega\gamma\gamma}=\dfrac{\alpha m_\omega^2}{384\sqrt{2}\pi^3m_e^2v_\chi}\epsilon_\chi^{\frac{2+2L_N}{L_\chi}}\Tr\left[\mathcal{Y}_\nu\mathcal{Y}_\nu^\dagger\right]\,.
\end{equation}

Tab. \ref{tab:exp-bounds} shows the numerical estimations for the predicted values of the Majoron coupling to photons, assuming that the trace gives an $\OO(1)$ number, as already discussed: the experimental bound is still far from the theoretical prediction. 

\begin{table}[h!]
\centering
\begin{tabular}{|c|c|c|c|}
\cline{2-4}
\multicolumn{1}{c|}{} &&&\\[-4mm]
\multicolumn{1}{c|}{}&$\lambda_{\omega\gamma\gamma}$&$\lambda_{\omega ee}$&$\lambda_{\omega\nu\nu}$\\[1mm]
\cline{1-4}
&&&\\[-4mm]
CASE NR1 &$[10^{-39},10^{-36}]\GeV^{-1}$&$[10^{-25},10^{-24}]$&\multirow{2}{*}{$[10^{-14},10^{-12}]$}\\[1mm]
\cline{1-3}
&&&\\[-4mm]
CASE NR2&$[10^{-34},10^{-32}]\GeV^{-1}$&$10^{-20}$&\\[1mm]
\hline
&&&\\[-4mm]
Exp. Upper bounds&$10^{-10}\GeV^{-1}$&$10^{-13}$&$10^{-5}$\\[1mm]
\hline
\end{tabular}
\caption{\em Predictions for the Majoron effective couplings to electrons, photons and neutrinos, for the window of the parameter space where Hubble tension is alleviated. In the last line, the corresponding experimental upper bounds.}
\label{tab:exp-bounds}
\end{table}

\subsection*{Coupling to electrons.} 
Astrophysical measurements can also constrain Majoron couplings. Ref.~\cite{Viaux:2013lha} provides an upper bound on the Majoron effective coupling to electrons 
\be
\LL_\omega^\text{low-energy} \supset i\, \lambda_{\omega ee}\,\omega\,\bar{e}\,e\,,
\ee
based on observations on Red Giants: 
\be
\lambda_{\omega ee}<4.3\times 10^{-13}\,.
\ee

The decay width of the Majoron to two electrons reads~\cite{Garcia-Cely:2017oco}
\be
\Gamma_{\omega ee}\simeq\dfrac{1}{8\pi}|\lambda_{\omega ee}|^2\, m_\omega\,,
\ee
where
\be
\lambda_{\omega ee}\simeq\dfrac{1}{8\pi^2}\frac{m_e}{v}\left(\left[\dfrac{m_Dm_D^\dagger}{vv_\chi}\right]_{11}-\dfrac12\Tr\left[\dfrac{m_Dm_D^\dagger}{vv_\chi}\right]\right)\,,
\ee
with $[\ldots]_{11}$ standing for the $(1,1)$ entry of the matrix in the brackets. Substituting explicitly the expression for $m_D$, the coupling becomes 
\begin{equation}
\lambda_{\omega ee}=\dfrac{1}{16\pi^2}\dfrac{m_e}{v_\chi}\epsilon_\chi^{\frac{2+2L_N}{L_\chi}}\left(\left[\mathcal{Y}_\nu\mathcal{Y}_\nu^\dagger\right]_{11}-\Tr\left[\mathcal{Y}_\nu\mathcal{Y}_\nu^\dagger\right]\right)\,.
\end{equation}
Assuming as before that the elements of the product $\mathcal{Y}_\nu\mathcal{Y}_\nu^\dagger$ are $\OO(1)$ numbers, also the Majoron-electron coupling is predicted to be much smaller than the corresponding experimental bound, as shown in Tab. \ref{tab:exp-bounds}. 

\subsection*{Coupling to neutrinos. Majoron emission in $0\nu\beta\beta$ decays.} 

The tree level couplings of the Majoron to neutrinos does not have an impact only on cosmology, but may be relevant for low-energy terrestrial experiments. In particular, searches for neutrinoless-double-beta decay could also be sensitive to processes in which Majorons are produced.

Current measurements set a lower bound on the half-life of the neutrinoless-double-beta decay of the order of $10^{26}$ years~\cite{KamLAND-Zen:2022tow,GERDA:2020xhi}.  However, these cannot be directly employed to constrain a process in which a Majoron is also produced, due to the different energy distribution of the emitted electrons. The NEMO-3 collaboration performed a dedicated search~\cite{Arnold:2018tmo}, setting a lower bound of the order of $10^{22}$ years on Majoron emission in $0\nu\beta\beta$ decay. This corresponds to a limit on the Majoron-neutrino coupling that reads
\be
\lambda_{\omega\nu\nu}<10^{-5}\,,
\ee
where $\lambda_{\omega\nu\nu}$ is defined in Eq.~\eqref{lambdaomeganunu}. Note that the exact constraint depends on the choice of nuclear matrix elements.

\begin{figure} [h!] \centering 
\resizebox{0.3\textwidth}{!}{\begin{tikzpicture}
\begin{feynman}
\vertex (vfi1){$d$};
\vertex [right=of vfi1](vi1);
\vertex [right=of vi1](fifow);
\vertex [right=of fifow] (vi2);
\vertex [right=of vi2] (vfo1){$u$};
\vertex [below=of fifow] (nuew) ;
\vertex [right=of nuew] (vi3);
\vertex [right=of vi3] (eo1) {$e$};
\vertex [below=of nuew] (nunum) ;
\vertex [right = of nunum] (vi4);
\vertex [right = of vi4] (mo) {$\omega$};
\vertex [below= of nunum] (nuew2);
\vertex [right= of nuew2] (vi5);
\vertex [right= of vi5] (eo2) {$e$};
\vertex [below= of nuew2] (fifow2);
\vertex [left= of fifow2] (vi6);
\vertex [left= of vi6] (vfi2) {$d$};
\vertex [right= of fifow2] (vi7);
\vertex [right= of vi7] (vfo2) {$u$};

\diagram* {
      (vfi1) -- [fermion] (fifow) -- [fermion] (vfo1),
      (vfi2) -- [fermion] (fifow2) -- [fermion] (vfo2),
      (fifow) -- [boson, edge label=$W^-$] (nuew),
      (fifow2) -- [boson, edge label'=$W^-$] (nuew2),
      (nuew) -- [fermion] (eo1),
      (nuew2) -- [fermion] (eo2),
      (nuew) -- [solid, edge label'=$\nu$] (nunum),
      (nuew2) -- [solid, edge label=$\nu$] (nunum),
      (nunum) -- [scalar] (mo),
    };
\end{feynman}
\end{tikzpicture}}
\caption{\em Feynman Diagram for the neutrinoless-double-beta decay with the emission of a Majoron.}
\label{fig:0nu2betaomega}
\end{figure}

The predicted value of the Majoron-neutrino coupling can be read out in Tab. \ref{tab:exp-bounds} and it is much smaller than the corresponding experimental value and the bound from {\tt Planck}~\cite{Escudero:2019gvw} ($\lambda_{\omega\nu\nu}<\OO(10^{-12})$).

\subsection*{Coupling with the SM Higgs. Higgs invisible decay.}

Besides the Majoron, also the radial component of $\chi$ is present in the spectrum and does play a role modifying the Higgs scalar potential. Indeed, the most general scalar potential containing $H$ and $\chi$ can be written as
\be
V(H,\chi)=-\mu^2H^\dag H + \lambda\left(H^\dag H\right)^2-\mu_\chi^2 \chi^*\chi+\lambda_\chi \left( \chi^*\chi\right)^2 + gH^\dag H\chi^*\chi
\ee
The minimisation of such potential leads to VEVs for the two fields that read
\be
v^2=\dfrac{4\lambda_\chi\mu^2-2g\mu_\chi^2}{4\lambda \lambda_\chi-g^2}\,,\qquad\qquad
v_\chi^2=\dfrac{4\lambda\mu_\chi^2-2g\mu^2}{4\lambda \lambda_\chi-g^2}\,.
\ee
The parameters of this scalar potential need to be such that $v$ takes the electroweak value and $v_\chi$ acquires the values in Tab.~\ref{tab:param}.

Due to the mixed quartic term, the two physical scalar fields $h$ and $\sigma$ mix in the broken phase and the mass matrix describing this mixing is given by
\be
\cM^2=
\left(
\begin{array}{cc}
2\,\lambda\, v^2& g\,v\,v_\chi \\
g\,v\,v_\chi& 2\,\lambda_\chi\, v_\chi^2 \\
\end{array}\right)\,.
\ee
The two eigenvalues that arise after diagonalising this mass matrix are the following
\be
M_{h,\sigma}^2=\lambda v^2+\lambda_\chi v^2_\chi\pm\left(\lambda v^2-\lambda_\chi v_\chi^2\right)\sqrt{1+
\tan^22\vartheta}\,,
\label{ScalarMasses}
\ee
where
\be
\tan2\vartheta=\dfrac{g\,v\,v_\chi}{\lambda_\chi v_\chi^2-\lambda v^2}\,.
\ee
The lightest mass in Eq.~\eqref{ScalarMasses} corresponds to the eigenstate mainly aligned with the SM Higgs, while the heaviest state is mainly composed of the radial component of $\chi$. Its mass can be as small as a few hundred GeV or much larger than the TeV. From the relation between the mixed quartic coupling $g$ and the physical parameters, 
\be
g=\dfrac{M_\sigma^2-M_h^2}{2\,v\,v_\chi}\sin2\vartheta\,,
\ee
it is possible to straightforwardly study the dependence of $M_\sigma$ with the other model parameters. 

The mixing of the Higgs with the new scalar also yields a coupling of the former to two Majorons. Expanding the kinetic term of $\chi$ yields a $\sigma\omega\omega$ term, that in turn produces a coupling $h\omega\omega$ after an insertion of the scalar mixing. Thus, a new decay channel for the Higgs opens up, contributing to its invisible width. The rate of this process is given by 
\begin{equation}
\Gamma_{h\to\omega\omega}=\dfrac{\sin^2{\vartheta}M_h^3}{32\pi v_\chi^2}\,.
\end{equation}
These effects on Higgs phenomenology have an impact on the signal strength $\mu_h$, defined as the ratio of observed Higgs events with respect to the SM expectation. This quantity has been measured to be in perfect agreement with the SM by both ATLAS and CMS~\cite{ATLAS:2022vkf, CMS:2022dwd}. In our setup, both the production cross section and the visible decay rates are modified. The former is suppressed by a factor $\cos^2{\vartheta}$, while the latter diminishes due to the appearance of the invisible channel. Explicitly,
\begin{equation}
    \mu_h= \cos^2{\vartheta}\left(1-\frac{\Gamma_{h\to\omega\omega}}{\Gamma_{h}}\right)\,,
\end{equation}
where $\Gamma_{h}$ is the total decay width of the Higgs boson. The combination of the ATLAS and CMS measurements yields a lower limit of $\mu_h>0.94$~\cite{Fernandez-Martinez:2022stj} at the 95\% CL, which translates into an upper bound on $\sin{\vartheta}$ that is proportional to $v_\chi$. 

\begin{figure}[b!]
\centering
\includegraphics[width=0.6\textwidth,keepaspectratio]{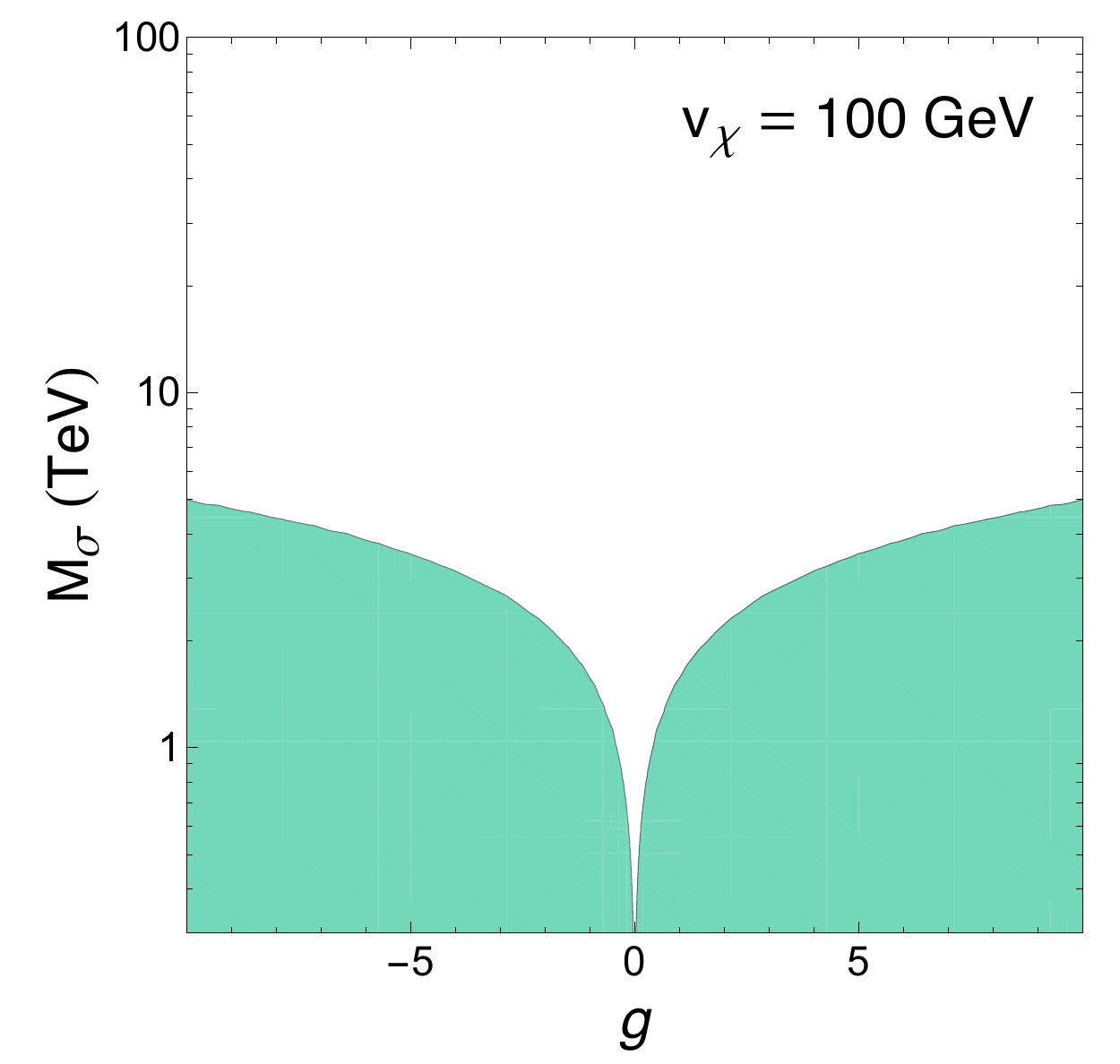}
\caption{Parameter space of the mass of the radial component of the new scalar as a function of its quartic coupling with the Higgs doublet. The green region is excluded by bounds on Higgs signal strength. We choose $v_\chi=100$ GeV, although the bounds are basically insensitive to this quantity.}
\label{Fig:SigmaMassPlots}
\end{figure}
These constraints can be portrayed in the parameter space composed of the mass of $\sigma$ and the quartic coupling $g$, as seen in Fig.~\ref{Fig:SigmaMassPlots}, in which the green areas are ruled out. Although we choose a particular value for $v_\chi$, we find that the bounds are basically independent on this quantity. Clearly, $M_\sigma$ can reach very large values without requiring any fine-tuning on $g$; conversely, $\sigma$ can only be light if $g$ is close to zero. Note that the upper bound on the Higgs invisible branching ratio (that currently stands at a 13\%~\cite{ParticleDataGroup:2022pth}) constitutes an independent source of limits, which exhibit a very similar dependence with the parameters of the model. However, these bounds are always looser than those given by signal strengths, so we do not include them in our results.
\subsection{Heavy Neutrinos}

In both cases discussed in Sect.~\ref{sec:Mechanism}, the heavy neutrino masses lie in ranges that may lead to detection in various experimental facilities. Neutrinos with masses ranging from tens to hundreds of MeV can be probed for and be potentially detected at beam dump or even near detectors of neutrino oscillation experiments~\cite{Gorbunov:2007ak,Atre:2009rg,Bondarenko:2018ptm,SHiP:2018xqw,Bondarenko:2019yob,Ballett:2019bgd,Berryman:2019dme,Coloma:2020lgy}, such as DUNE or SHiP, whereas those with masses in the range of tens to hundreds of GeV have interesting prospects of being produced at the LHC or future colliders~\cite{delAguila:2008cj,Atre:2009rg,Antusch:2015mia,Deppisch:2015qwa,Antusch:2016ejd,Cai:2017mow,Dev:2018kpa,Pascoli:2018heg}. 

On the other hand, given their extremely small couplings, the heavy neutrinos produced in the early Universe would not be Boltzmann suppressed when decoupled from the thermal bath, leading to an unacceptably large contribution to the relativistic degrees of freedom of the Universe after their subsequent decay~\cite{Sato:1977ye,Gunn:1978gr,Hernandez:2013lza,Hernandez:2014fha,Vincent:2014rja}. If their decay takes place before the onset of BBN, the decay products would quickly thermalise and BBN would then proceed as in the standard $\Lambda$CDM scenario. However, if the decay of the heavy neutrinos happens after BBN and neutrino decoupling, their contribution to the effective number of neutrinos would be too high and ruled out. If the decay takes place during BBN, the decay products could also alter the production of primordial helium and strong constraints also apply~\cite{Dolgov:2000pj,Ruchayskiy:2012si,Gelmini:2020ekg,Boyarsky:2020dzc}. This would be the situation of CASE NR1, for which the neutrino masses and mixings predict decay rates comparable to or larger than the onset of BBN. Conversely, the larger masses that characterize CASE NR2 lead to decays faster than BBN, eluding these cosmological constraints. Hence, BBN and CMB observations disfavor CASE NR1 unless the heavy neutrino decay is faster than BBN in some part of the parameter space or some other modification of the standard $\Lambda$CDM scenario is considered. Indeed, if the heavy neutrinos decay after BBN, for heavy neutrino masses in the range $[3.5,200] \MeV$, then the bound on the mixing is~\cite{Vincent:2014rja}:
\be
\sin^2\theta_s\equiv \dfrac{\langle m_\nu\rangle}{\langle M_N\rangle}\lesssim10^{-15}-10^{-17}\,,
\ee
much smaller than the expected value that can be read in Tab.~\ref{tab:HeavyNetrinos}.

\begin{table}[h!]
\centering
\begin{tabular}{|c||c|c||c|c|}
\cline{2-5}
\multicolumn{1}{c|}{}	&&&&\\[-4mm]
\multicolumn{1}{c|}{}	& $\langle M_N\rangle$ & $\sin^2\theta_s$ & $\Gamma_{N\to3\nu}^Z$ & $\Gamma_{N\to3\nu}^\omega$\\[1mm] \hline
&&&&\\[-4mm]
CASE NR1 	& $[3.5,200] \MeV$ & $[2.5\times10^{-10},\,1.4\times 10^{-8}]$ & $\OO(10^{-38})$ & $\OO(10^{-68})$\\[1mm] \hline
&&&&\\[-4mm]
CASE NR2 	& $[35.4,707] \GeV$ & $[7.1\times10^{-14},\,1.4\times 10^{-12}]$ & $\OO(10^{-27})$ & $\OO(10^{-66})$ \\[1mm]
\hline
\end{tabular}
\caption{\em Expectation for the heavy neutrino mass and mixing between heavy and active neutrinos.}
\label{tab:HeavyNetrinos}
\end{table}

%%%%%%%%%%%%%%%%%%%%%%%%%%%%%%%%%%%%%%%%%%%%%%%%%%%%%%%%%%%%
\section{Conclusions}
%%%%%%%%%%%%%%%%%%%%%%%%%%%%%%%%%%%%%%%%%%%%%%%%%%%%%%%%%%%%
\label{sec:concl}

The Hubble tension may well be due to systematic uncertainties but it might point to a deviation from the $\Lambda$CDM model or an extension of the Standard Model of particle physics. In particular, the presence of a Majoron with couplings to the active neutrinos represents an interesting avenue to alleviate this tension. This paper is focussed on a mechanism where neutrino masses and the presence of a Majoron are explained by the spontaneous breaking of Lepton Number. Besides the introduction of three RH neutrinos and a new complex scalar singlet, three sets of Lepton Number charge assignments are identified that correctly explain the lightness of the active neutrinos and provide the Majoron with couplings that lower the Hubble tension from $4.4\sigma$ down to $2.5\sigma$.

This mechanism is completely general and could be embedded in different types of flavour models. A compelling case presented here is the embedding in the Minimal (Lepton) Flavour Violating setup:  Lepton Number is part of the complete flavour symmetry and therefore naturally arises in this context. The non-Abelian factors are mainly responsible of the intergenerational mass hierarchies and of the mixings, while  Lepton Number, besides being associated to the Majoron, is involved in the explanation of the smallness of the active neutrino masses with respect to the top mass. The ratios between the bottom and $\tau$ masses can be explained via the spontaneous breaking of the PQ symmetry, which completes the non-Abelian flavour symmetry together with Lepton Number. A QCD axion arises as a byproduct of the PQ symmetry breaking, allowing to solve the Strong CP problem within the same framework. 

The Majoron only couples at tree level to neutrinos. The neutrinoless-double-beta decay represents a natural observable where to look for it, but the model prediction turns out to be much smaller than the current experimental sensitivity and much stronger constraints can be derived from its impact in neutrino free streaming and the effective number of neutrino species from CMB observations. Effective couplings to charged fermions and photons arise at one and two loops, respectively. For this reason, the model is not significantly constraint from CAST and Red Giant observations that otherwise provide stringent bounds for light pseudoscalars.

Due to the mixing between the physical Higgs and the radial degree of freedom of the scalar field producing the Majoron, the latter does couple to the Higgs. This has an impact on the invisible Higgs decay, that indeed provides a strong bound on the VEV of the new scalar, that is, on the Lepton Number breaking scale. However, a natural choice in the model is to assume the radial component to be sufficiently heavy to be integrated out, and correspondingly its mixing with the Higgs gets suppressed, relaxing the bound on the VEV.

Finally, heavy neutrinos are expected to be much lighter than in the traditional type-I Seesaw scenario, in the MeV or GeV ranges depending on the specific case considered. In these mass ranges the decay products of the neutrinos could significantly affect the evolution of the early Universe. Indeed, for masses in the MeV range, decays during Big Bang Nucleosynthesis could alter the production of primordial helium while later decays would imply too large an injection of relativistic species. Conversely, for neutrino masses above the GeV, their decays are typically faster than Big Bang Nucleosynthesis and the decay products thermalise without altering the rest of the thermal history of the Universe. This option is thus preferred with respect to the lighter masses.

The result is a model where fermion masses and mixings can be correctly described, protecting the flavour sector from large deviations from the standard predictions with a new physics scale of the order of a few TeV, solving the Strong CP problem and softening at the same time the Hubble tension. This model could be tested indirectly at colliders looking at the invisible Higgs decay and searching for relatively light heavy neutrinos or for the radial component of the scalar field that generates the Majoron.

%%%%%
%%%%%%%%%%%%%%%%%%%%%%%%%  Acknowledgments   %%%%%%%%%%%%%%%%%%%%
%%%%%s
\section*{Acknowledgements}

The authors acknowledge partial financial support by the Spanish MINECO through the Centro de excelencia Severo Ochoa Program under grant SEV-2016-0597, by the Spanish ``Agencia Estatal de Investigaci\'on''(AEI) and the EU ``Fondo Europeo de Desarrollo Regional'' (FEDER) through the projects FPA2016-78645-P and PID2019-108892RB-I00/AEI/10.13039/501100011033. This project has received support from the European Union's Horizon 2020 research and innovation programme under the Marie Sk\l odowska-Curie grant agreement No 860881-HIDDeN. L.M. acknowledges partial financial support by the Spanish MINECO through the ``Ram\'on y Cajal'' programme (RYC-2015-17173).

%%%%%
%%%%%%%%%%%%%%%%%%%%%%%%%  Bibliography    %%%%%%%%%%%%%%%%%%%%%%%%
%%%%%

\footnotesize

%
%\bibliography{biblio}{}
\bibliographystyle{BiblioStyle}

\providecommand{\href}[2]{#2}\begingroup\raggedright\endgroup

\end{document}